# Development of Microcontroller Based Smart Grid Framework

Saurav Biswas

*Department Electrical and Electronic Engineering (EEE), Bangladesh University of Business and Technology*
*Bangladesh*

*Abstract— Smart grid technology has been recognized as a promising solution for the next-generation energy efficient electric power systems to mitigate energy crisis. Smart grid provides highly consistent and reliable services, efficient energy management practices, smart metering integration, automation and precision decision support systems and self-healing facilities. The smart power grid introduces a sensing, monitoring, and control system that provides end users with the cost of energy at any moment through real-time pricing. The classical power system operation has no control over the loads except in an emergency situation when a portion of the loads can be dropped as needed to balance the power grid generation with its loads. Furthermore, the smart power grid supplies the platform for the use of renewable energy sources that acts as a safeguard against a complete blackout of the interconnected power grid. In this work, all the concepts involved in smart grid mechanism is implemented with PIC18F452 microcontroller and other supplementary components. A solar module with storage capacity is connected to the proposed system for minimizing grid energy consumption and plays a primary energy source for maximum utilization of green energy. This intelligent device may inherently reduce the consumption of electrical energy during peak hours and allow consumers to sell back electricity into grid using bi-directional technique. Barriers, challenges, benefits and future trends regarding the technologies and the role of users have also been discussed in this paper.*

## I. INTRODUCTION

The utility sector has always taken advantage of advances in communication and information technology to improve efficiency, dependability, security, and service quality. Increasing complexity in managing the bulk power grid, growing environmental concerns, energy sustainability and independence, demand growth, and the pursuit of service quality all highlight the need for a quantum leap in use of such technology. This leap toward a "smarter" grid is widely referred to as "smart grid." [1]. The operational data acquired by the smart grid and its sub-systems will allow system operators to quickly identify the optimal plan for securing against various situations like as assaults, vulnerabilities, and so on. A fixed price is charged to energy users in a traditional power grid. The daily peak load operation, on the other hand, has the largest energy cost. Except in an emergency, when a portion of the loads can be reduced as needed to balance the power grid generation with its loads, traditional power system operation has no control over the loads. As a result, most grid elements are only employed for a brief time during peak power demand and are idle during normal operation.

To maintain reliability, stability, and efficiency, today's electric grid was built as a vertical system that included generation, transmission, and distribution, as well as controls and devices. However, system operators are now confronted with additional problems, such as RER penetration in legacy systems, fast technological development, and a variety of market competitors and end users. The smart grid, the next generation, will include communication support schemes and real-time measuring techniques to improve resiliency and forecasting, as well as guard against internal threats. The smart grid's design framework is focused on unbundling and reorganizing the electricity industry, as well as optimizing its assets. Significant infrastructure investment in the form of a communication system, cyber network, sensors, and smart meters must be implemented to curb system peak demands when the cost of electric energy is highest for an effective smart power grid system design and operation. The smart power grid includes a sensing, monitoring, and control system that provides end-users with real-time energy cost pricing. Furthermore, smart metering's sophisticated control systems allow energy customers to react quickly to real-time pricing. It safeguards the connected electrical networks from total blackouts induced by man-made or natural calamities. It also allows for the disaggregation of the connected electrical grid into smaller regional clusters. Furthermore, the smart power grid allows anybody to become an energy producer by allowing anyone to employ solar or wind energy, fuel cells, or combined heat and power (CHP) energy sources, as well as participate in the energy market by buying or selling energy via the smart meter connection [2].

In next generation electrical systems that combine different renewable energy sources, automated and smart management is a critical component for evaluating the efficacy and efficiency of these power systems [3]. The Smart Grid name is meant for management automation and intelligence in terms of a variety of benefits over existing systems in terms of digitalization, flexibility, intelligence, resilience, sustainability, and standards [4-6]. Intelligent control centres should be able to remotely monitor and engage with electrical equipment in real time, intelligent transmission infrastructures should employ new technologies to enhance power quality, and intelligent substations should consciously coordinate their local devices [7-8]. Thanks to significant advancements in system automation and intelligence, the notion of the Energy internet [9] has been proposed as having great possibilities for the future energy usage paradigm spanning all phases of power generation, storage, transmission, and distribution.

The world is ushering in a new era, which is recognized by technology enthusiasts, with the advent of cutting-edge technology including such cellular communications [10-13], internet of things [14], [15], antenna design [16]-[17], sensor design [18]-[21], and advanced optics [22]-[24]. For a variety of reasons, harvesting from locally available renewable energy sources (RES) is a popular approach that is promoted as an established technology [25]-[32]. For starters, it reduces reliance on fossil fuels. Second, renewable energy sources are abundant all around the globe. Third, by decreasing the cost of generating electricity and lowering the carbon content, a significant quantity of green energy may be produced from renewable sources of energy. Several studies [33]-[37] have been carried out in an attempt to develop a long-term, dependable, and energy-efficient supply network based on locally accessible renewable energy sources.

The deployment of sophisticated metering infrastructure offers energy end customers with real-time pricing. Parallel to this, the increasing use of renewable energy sources is creating a foundation for autonomous or local control of microgrids connected to the local power grid. Fault detection, isolation, and restoration will be provided through a distributed autonomous control system. The autonomous control and real-time pricing also improve feeder voltage efficiency, lowering feeder losses and lowering plug-in electric car feeder peak demand. Community energy storage will become another essential aspect for microgrid control as storage technology matures, allowing the energy user to become an energy producer. To realize the benefits of a smart power grid, these interconnected technologies require a coordinated modelling, simulation, and analytic system.

All of the principals involved in the smart grid mechanism are realized in this work using a PIC16F887 microcontroller and other auxiliary components. A PV module with storage capability is linked to the proposed system to reduce grid electricity usage even while serving as a main energy source to maximize clean energy use. This clever gadget may minimize electrical energy usage during peak hours and allow consumers to sell electricity back into the grid via a bi-directional approach. This article also discusses technology barriers, problems, rewards, and future developments, as well as the role of users. The rest of the paper is as follows: section II discusses the overall system model. The result and discussion are explained in section III and section IV concludes the paper.

## II. SYSTEM MODEL

This section will discuss the overall system model including the structure of the smart grid, the microcontroller, solar PV module, Phasor Measurement Units (PMU), solar inverter, and smart energy meter. A renewable energy based smart grid (MG) is depicted in Figure 1. PV, wind power, and green energy sources such as fuel cells and high-speed micro turbine generating stations are all part of the MG systems. GS's are designed to either supply electricity to the local grid or import power based on pre-arranged contracts.

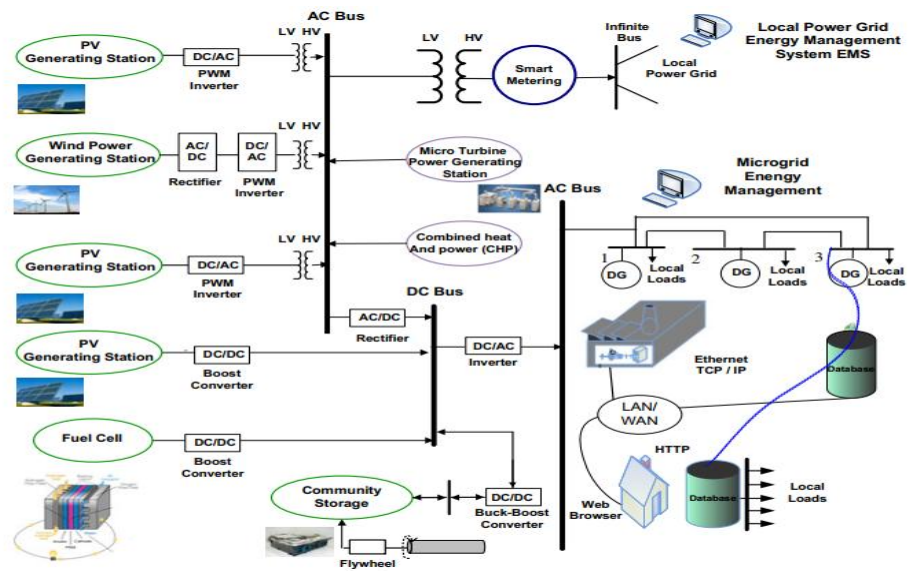

Figure 1. Basic structure of smart grid.

The AC bus of SG systems is directly connected to combined heat and power (CHP) and micro turbine producing facilities, as well as certain induction generator-based wind generation. PV, variable-speed wind, and fuel cell producing plants, on the other hand, produced DC electricity. PV electricity may be directly linked to SG systems' AC bus using DC-AC inverters. The use of a DC-DC boost converter to step up the DC voltage and higher-rated DC-AC inverters is an alternative design. When MRG systems are implemented, the AC bus voltage and frequency are synced to the AC bus of the local power grids the energy management of the local power grid controls frequency [38-42]. However, once MRG systems are disconnected from the local power grid, MRG's energy management system must maintain control over the MRG AC bus voltage and frequency to ensure stable operation. When the MRG is disconnected from the local power grid, load control is critical for maintaining adequate load and generation balance. Figures. 2 and 3 present the block diagram of the microcontroller based smart grid system and the internal circuit of the proposed model respectively.

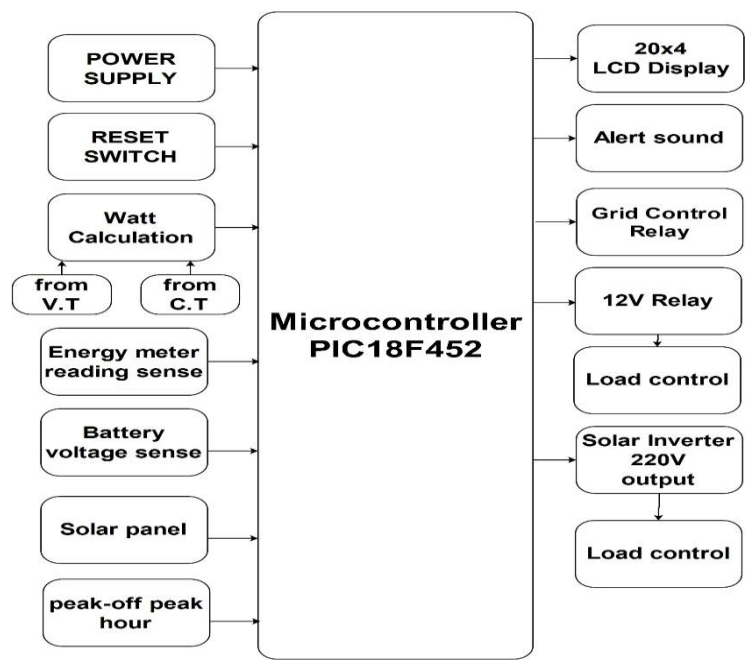

Figure 2. Block diagram of the proposed model.

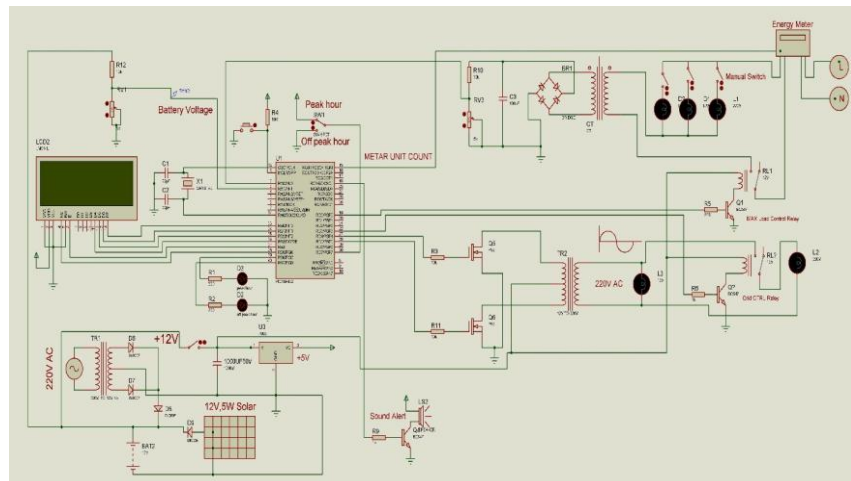

Figure 3. Circuit diagram of the prposed model.

The circuit schematic for this project is shown in Figure 3.3.1 above. The PIC18F452 IC, a 40DIP microcontroller with 35 I/Os, is the primary component. This system is clever in that it detects only a few situations. Our major priority setting is peak and off peak hours. With the pin no 30 of the microcontroller, we fix a button to alter the condition of peak hour and off peak hour. We will be able to determine how much power is utilized as well as the current operational load using this system. We attached a single phase energy meter to pin 15 of the microcontroller and got a pulse. This system was intended to handle a maximum load of 200 watts. We constructed an inverter using P55 mosfet switching oscillated from microcontroller pins 27 and 28 to convert solar pv dc electricity to 220 V 50Hz AC. One 12V battery is used to store solar pv generation, and its charging and discharging rates are controlled by the microcontroller pin 3 as planned. Other auxiliary components, such as the LCD display, are linked to pins 33 to 38 [43-49].

### A. Microcontroller (PIC18F452)

This 10 MIPS (100 nanosecond instruction execution) CMOS FLASH-based 8-bit microcontroller packs Microchip's powerful PIC® architecture into a 40- or 44-pin package and is upwards compatible with the PIC16C5X, PIC12CXXX, PIC16CXX, and PIC17CXX devices, providing a seamless migration path of software code to higher levels of hardware integration. The PIC18F452 has a C compiler-friendly development environment, 256 bytes of EEPROM, Self-programming, an ICD, 2 capture/compare/PWM functions, 8 channels of 10-bit Analog-to-Digital (A/D) converter, synchronous serial port that can be configured as either 3-wire Serial Peripheral Interface (SPI) or 2-wire Inter-Integrated Circuit (I2CTM) bus, and Addressable Universal Asynchronous Receiver Transmitter (AUSART) [50-56].

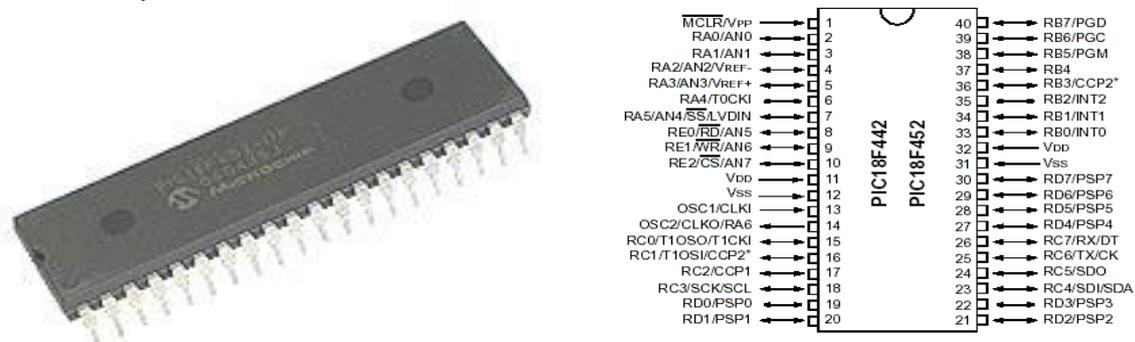

Figure 3. Microcontroller PIC18F452 Pin configuration

It's suitable for manufacturing equipment, instrumentation and monitoring, data collecting, power conditioning, environmental monitoring, telecommunications, and consumer audio/video applications because of all of these qualities.

***B. Solar PV Module***

Solar is the direct form of electricity at the atomic level. Some materials have a photoelectric effect, which allows them to absorb photons of light and release electrons. When these free electrons are gathered, an electric current is generated, which may be used to create electricity [57-58].

***C. Phasor Measurement Units (PMU)***

Synchro phasors or Phasor Measurement Units provide operators with a time-stamped picture of the power system. The PMUs include information such as locations and other network characteristics, as well as bus voltage and branch current phasors. At the same time, phasor measurements are acquired with great accuracy from multiple places of the power system, allowing an operator to see exact angular difference between different locations. GPS receivers are built into PMUs, allowing for the synchronization of readings collected at different locations. The PMU module is integrated with other current functions as an additional feature in microprocessor-based instrumentation such as protection relays and Disturbance Fault Recorders (DFRs). The technique for transmitting PMU data to the Phasor Data Concentrator is specified in the IEEE standard [59-64].

***D. Solar inverter***

The variable direct current (DC) output of a photovoltaic (PV) solar panel is converted into a utility frequency alternating current (AC) that may be supplied into a commercial electrical grid or utilized by a local, off-grid electrical network by a solar inverter, also known as a PV inverter. It is a critical balance of system (BOS) component in a photovoltaic system that permits the use of conventional AC-powered equipment. Solar power converters have developed features such as maximum power points, tracking, and anti-islanding protection for use with photovoltaic arrays [65-68].

***E. Energy Meter***

An electricity meter, electric meter, electrical meter, or energy meter is a device that monitors the amount of electric energy used by a household, a business, or an electrically powered object [69].

### III. RESULTS

This section provides the research findings as well as a discussion of the work. We've also spoken about the benefits, drawbacks, and limitations of the present iteration of the protection system.

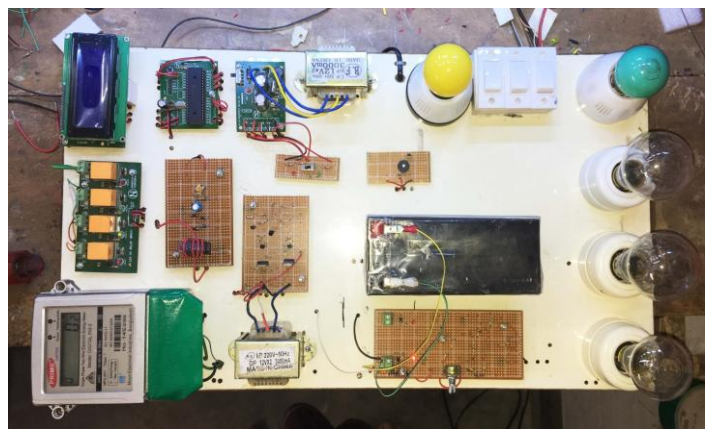

Figure 4. Experimental setup

The goal of this project was to create an intelligent and clever energy management system that could work in a variety of situations. There are two load situations, peak hour and off peak hour, which result in differing unit prices. The charging and discharging mechanism from solar panels is controlled by a microprocessor, and when the battery voltage remains below 12V, it will not invert to AC and hence will not connect to the grid. When the battery voltage maintains over 12 volts, it will only connect to the grid during off-peak loads. Aside from that, the system manages the maximum operational load as planned. As a result, the difficulty was to ensure that all of the requirements were met while working under the supervision of a single unit.

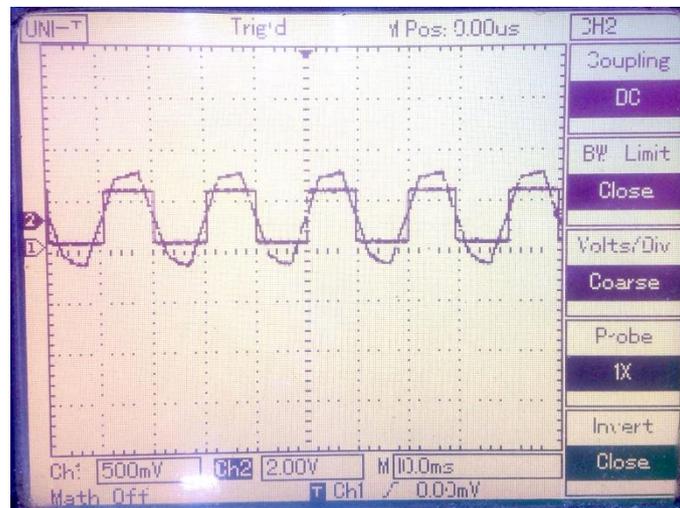

Figure.1 Output waveform of utility grid and generated ac signal

This entire system operates in accordance with a set of criteria or standards that we devised. It serves two primary purposes. The load side is one, while the power supply is the other. Because this is a dual-source power management system, with a maximum operational load of 200 Watts, the utility grid will take priority. We set up a battery storage system to store and discharge dc electricity as needed. During peak hours, when the user's usage reaches 200 W, the system will emit a beep to reduce the load for a few seconds before disconnecting the user from the utility. To store and discharge dc power as needed, we put up a battery storage system. When the user's consumption surpasses 200 W during peak hours, the system will emit a beep to lower the load for a few seconds before disconnecting the user from the utility. It is suited for peak hour and power down conditions since it has both an on-grid and a battery backup mechanism. The zero-crossing detection method of synchronization ensures a safe and smooth start. We can regulate and maintain the amount of charge every month for ourselves because this is an intelligent system, and we can also utilize off-grid regions. On the other hand, Unless the utility frequency fails to synchronize, the protection mechanism was not considered, and the utility frequency should be 50Hz.

## IV. Conclusions

Recent environmental awareness as a result of conventional power plants has sparked interest in contemporary smart grid technology and its integration with climate-friendly green renewable energy. Smart grid operations enable for higher penetration of variable energy sources by allowing for more flexible system management. Experiments were conducted out to look at the effects of renewable energy sources in a smart power network, particularly large-scale PV adoption. Voltage fluctuations and harmonic injection rise as PV penetration grows, according to experimental and modeling results. Several initiatives and strategies are being implemented to alleviate the energy deficit; nevertheless, as demand grows, the government has set a goal of reaching 26000MW generating capacity by 2021. Solar PV systems will be one of the most important renewable energy sources for us in order to meet this goal. However, due to a lack of technical devices and reliance on foreign technology, the cost of per watt generation remains higher than in other developed countries; the government has launched several programs to subsidize this sector; we hope that our approach of developing a "Smart Grid System" can help to alleviate this situation.